\documentclass[aps,prl,twocolumn,amsmath,amssymb]{revtex4-2}

\usepackage{graphicx}
\usepackage{color}
\usepackage[hidelinks]{hyperref}
\usepackage{orcidlink}

\hypersetup{colorlinks=true, linkcolor=black, citecolor=black, urlcolor=blue}

\begin{document}

\title{Temperature-Enhanced Critical Quantum Metrology}

\author{Laurin Ostermann\,\orcidlink{0000-0001-8508-9785}}
\author{Karol Gietka\,\orcidlink{0000-0001-7700-3208}}
\email[]{karol.gietka@uibk.ac.at}

\affiliation{Institut f\"ur Theoretische Physik, Universit\"at Innsbruck\\Technikerstra{\ss}e\,21a, A-6020 Innsbruck, Austria} 

\begin{abstract}
We show that the performance of critical quantum metrology protocols, counter-intuitively, can be enhanced by finite temperature. We consider a toy-model squeezing Hamiltonian, the Lipkin-Meshkov-Glick model and the paradigmatic Ising model. We show that the temperature enhancement of the quantum Fisher information can be achieved by adiabatic preparation of the state close to the critical point and by preparing it directly in the proximity of the critical point. We also find a relatively simple, however, non-optimal measurement capable of harnessing finite temperature to increase the parameter estimation sensitivity. Therefore, we argue that temperature can be considered as a resource in critical quantum metrology.
\end{abstract}
\date{\today}
\maketitle

\emph{Introduction}---Quantum metrology~\cite{QM2006Llyod,QM2011Lloyd} is one of the most promising directions in quantum techmologies~\cite{QT2003Milburn,paris2009qefqt,Rabl2015QT,QT2018Acin}. It aims at exploiting quantum phenomena in order to enhance measurement resolution and sensitivity of physical parameters. This is typically achieved by preparing non-classical states~\cite{pirandola2018rmp,QM2018rmp,QM2020Sciarrino}, such as squeezed states~\cite{wineland1992squeezing,MA201189spins}, and subsequently exposing them to a perturbation related to an unknown parameter. One prominent example is Ramsey interferometry~\cite{schmiedmayer2018ramsey,Schulte2020ramsey,kaubruegger2021ramsey} where the perturbation can be related to magnetic field splitting the atomic energy levels or to a laser field exciting the atoms. By performing an appropriate measurement on the final (perturbed) state it is possible to estimate  the unknown parameter from the measurement outcomes~\cite{helstrom1969detection}. Repeated measurements result in a histogram, where its average value is the estimated unknown parameter and its width is related to the uncertainty of the estimation: the narrower the histogram the better determined the unknown parameter. The minimal width of the histogram is related to the inverse of the quantum Fisher information through the Cram\'er-Rao bound~\cite{helstrom1969detection,caves1994statisticaldistance}. Therefore, the Quantum Fisher information plays an essential role in quantum metrology and increasing its value can be seen as one of the central aims of the field.

The simplest way to increase the quantum Fisher information is to increase the number of elements, e.g\ atoms or photons, constituting the system. Intuitively, this can be understood as equivalent to performing many measurements simultaneously. In this case the quantum Fisher information scales linearly with the number of elements $N$, which is called the standard quantum (or shot noise) limit. Another simple way of increasing the quantum Fisher information is to expose the initial state to the perturbation for a longer time $t$. With this, one can imprint more information about an unknown parameter on the initial state. Consequently, the quantum Fisher information (usually) scales quadratically with time~\cite{pang2014generalHparameter,pang2017timedepmetrology}. A more challenging way of increasing the quantum Fisher information is to use non-classical correlations~\cite{smerzi2009entHL}, which can be understood as reducing the inherent quantum noise or fluctuations. Using maximally correlated (entangled) states allows for elevating quantum Fisher information scaling from linear with the number of constituents to quadratic, which is dubbed Heisenberg scaling. In the optimal scenario, the quantum Fisher information becomes $\mathcal{I} = N^2t^2$ which is the Heisenberg limit. Although introducing quantum correlations in the initial state can dramatically increase quantum Fisher information, it can also significantly complicate the entire protocol, especially the measurement stage. Due to the correlations, the possible set of measurement outcomes can grow tremendously~\cite{GUHNE20091}; thus, increasing the time of estimation, gathering and analyzing data in particular, which quantum Fisher information cannot quantify. Moreover, the quantum Fisher information assumes that an optimal measurement is performed on the perturbed state which sometimes may be unrealistic due to practical limitations. 

Although a number of techniques have been developed to counteract these issues~\cite{ostermann2013protected,smith2016approachingHL,dur2016largeQFI,szigeti2017interactionbased, colombo2022satin}, in any realistic case finite temperature and decoherence will deteriorate the quantum correlations and external noise will perturb the measurement. As a consequence, it is virtually impossible to attain the Heisenberg limit for a macroscopic quantum system~\cite{demko2012elusive}. Therefore, instead of trying to reach the Heisenberg limit, one can focus on devising practical and robust metrology protocols that overcome the standard quantum limit and ideally feature Heisenberg scaling. Note that quantum Fisher information can exhibit Heisenberg scaling while still being lower than the standard quantum limit. One example of such an approach relies on creating the correlations concurrently with storing the information about the unknown parameter~\cite{Hayes_2018concurrent,garbe2020criticalmetrology}. Excellent candidates for this approach are systems exhibiting phase transitions. At the critical point of a phase transition, the eigenstates of the system are extremely sensitive  to any changes of the system parameters~\cite{zanardi2007criticalscaling,zanardi2008informationgeometry,zanardi2008criticalityresource,mihailescu2023multiparameter}. If the final state of such a protocol is the ground state of the system, it will be immune to dissipation as the ground state is the lowest energy state. 

In this manuscript, we show that critical metrology is not only robust with respect to finite temperature but can even benefit from it. To this end, we demonstrate how finite temperature can increase the quantum Fisher information for a squeezing Hamiltonian, vastly studied in the literature in the context of critical metrology, for the Lipkin-Meshkov-Glick model, and for the paradigmatic Ising model. Furthermore, we present a simple measurement scheme for which the classical Fisher information at finite temperature is larger than the quantum Fisher information at $T=0$. Therefore, we argue that finite temperature can be considered as a resource in critical quantum metrology.

\emph{Toy Model: Squeezing Hamiltonian}---In order to understand how temperature can increase the quantum Fisher information we will first consider the toy model squeezing Hamiltonian ($\hbar = 1$)
\begin{align}\label{eq:toymodel}
    \hat H = \omega \hat a^\dagger \hat a - \frac{g}{4}\left(\hat a^\dagger + \hat a\right)^2.
\end{align}
The above Hamiltonian can be used to describe a phase transition in the quantum Rabi model, the Lipkin-Meshkov-Glick model and the Dicke model. It has been widely studied in the context of critical metrology~\cite{paris2014cqmLMG,garbe2020criticalmetrology,dynamicCQM2021JIanming,Gietka2021adiabaticcritical,Gietka2022understanding,plenio2002PRX,hotter2023combining}. By introducing abstract position and momentum operators the Hamiltonian~\eqref{eq:toymodel} can be mapped to a harmonic oscillator with an interaction-dependent effective frequency $\tilde \omega(g)=\omega\sqrt{1-g/g_c}$. For the critical coupling strength $g = g_c = \omega$, the harmonic potential vanishes and the eigenstates are infinitely squeezed Fock states of the non-interacting system. In general, the eigenstates are found to be
\begin{align}
    |n_\xi \rangle = \hat S(\xi)|n\rangle,
\end{align}
where $\hat S(\xi) = \exp\{\frac{1}{2}\left(\xi^*\hat a^2-\xi\hat a^{\dagger2}\right)\}$ is the squeeze operator and $\xi = \frac{1}{4} \ln\{1-g/g_c\}$ is the squeezing parameter. Using the definition of the quantum Fisher information $\mathcal{I}_\omega = 4\left[\langle \partial_\omega \psi| \partial_\omega \psi\rangle -|\langle \psi|\partial_\omega \psi \rangle|^2 \right]$, it is straightforward to show that for the $n$-th eigenstate (we assume $\omega$ is unknown)
\begin{align}\label{eq:qfi0}
    \mathcal{I}_\omega = 2[\partial_\omega \xi]^2(n^2+n+1),
\end{align}
where $\partial_\omega \xi = \frac{g}{4 \omega ^2 \left(1-{g}/{\omega }\right)}$. From the above expression, we see that { the ground state ($n=0$) sets the lower bound on the quantum Fisher information, while excited states increase the quantum Fisher information.} In a realistic scenario, however, where the system cannot be isolated from the environment, dissipation will quickly take away energy from the excited state and any enhancement deriving from $n$ will vanish. One way of forcing the system to robustly occupy excited states is to use fermions. In this case, the ground state is an antisymmetric superposition of single-particle eigenstates. In Ref.~\cite{gietka2023socmet} it was shown that for $N$ fermions occupying levels of the Hamiltonian \eqref{eq:toymodel}, the quantum Fisher information exhibits Heisenberg scaling with respect to $N$ for the ground state. A less obvious way that leads to a robust occupation of excited states (although incoherent) and thus an increase of the quantum Fisher information is using a finite temperature. This is rather counter-intuitive, therefore we will show how it comes about step by step.

For a finite temperature, the state of the system is in a mixed state described by a density operator (statistical mixture)
    $\hat \rho = \sum_{n} p_n |n\rangle \langle n|$,
where $n$ labels the excited states and $p_n$ is given by $p_n = \frac{1}{Z}\exp(- \beta E_n)$, $Z = \sum_n \exp(- \beta E_n)$ is the partition function (statistical sum) with $E_n$ being the energy of the $n$th excited state, and $\beta = 1/k_B T$ where $k_B$ is the Boltzmann constant and $T$ denotes temperature. In order to calculate the quantum Fisher information for a statistical mixture, we employ~\cite{bruansteincaves1994qfi}
\begin{align}\label{eq:qfiDM}
    \mathcal{I}_\omega =  \sum_n \frac{(\partial_\omega p_n)^2}{p_n} + 2 \sum_{n,m}\frac{(p_n-p_m)^2}{p_n+p_m}|\langle n| \partial_\omega | m \rangle |^2,
\end{align}
where $n$ ($m$) labels the eigenstates of the system. The two terms in Eq.~\eqref{eq:qfiDM} are usually referred to as the classical and the quantum contribution to the quantum Fisher information~\cite{paris2014cqmLMG}. Note that in typical approaches to quantum metrology, where the state is prepared first and subsequently exposed to the effect of an unknown perturbation, {the eigenvalues remain unaffected} and the probabilities do not carry any information about the perturbation. Hence the  first term in Eq.~\eqref{eq:qfiDM} can be neglected. Conversely, in quantum thermometry~\cite{sanpera2015indoptT,DePasquale2018quantumthermo,Mehboudi_2019sanperaReview,mossy2020tempfermi,correa2021globalthermor,mhboudi2022thermometrylimits,srivastava2023topological,yu2023criticalityenhanced,campbell2023thermodstrongcorrferm}, the information about the (unknown) temperature is stored in the probability distribution only.

The derivative $\partial_\omega$  with respect to the unknown parameter acting on the eigenstates of the Hamiltonian~\eqref{eq:toymodel} yields ($\xi$ is assumed real)
\begin{align}
    \partial_\omega |n_\xi \rangle = \frac{\partial_\omega \xi }{2}\left(\hat a^2-\hat a^{\dagger2}\right) |n_\xi\rangle.
\end{align}
The only non-zero contributions to the quantum Fisher information will originate from the overlap of states that differ by two excitations. Therefore, the quantum Fisher information can be rewritten as
\begin{align}
   \left[\partial_\omega \xi\right]^2 \sum_{n=0}\frac{(p_{n+2}-p_n)^2}{p_{n+2}+p_n}  (n+1)(n+2).
\end{align}
By inserting the probabilities with $E_n =  (n+1) \tilde\omega$, we obtain
\begin{align}\label{eq:toymodelqfi1}
   2 \left[\partial_\omega \xi\right]^2  \frac{\tanh (\beta  \tilde\omega )}{ \tanh \left(\frac{\beta \tilde \omega }{2}\right)}  \approx  4 \left[\partial_\omega \xi\right]^2  ,
\end{align}
where the approximation is valid for $k_B T  \gg \tilde\omega$. This expression is always greater than the quantum Fisher information from Eq.~\eqref{eq:qfi0} {with $n=0$ (ground state)}. The contribution from the probabilities can be easily calculated to be
\begin{align}
\label{eq:toymodelqfi2}
   \sum_n \frac{(\partial_\omega p_n)^2}{p_n} &=  \frac{\beta ^2 \left(2-\frac{g}{\omega }\right)^2 \text{csch}^2\left(\frac{1}{2} \beta \tilde \omega  \right)}{16 \left(1-\frac{g}{\omega }\right)} \nonumber\\ & \approx \frac{\left(2-\frac{g}{\omega }\right)^2}{4 \omega ^2 \left(1-\frac{g}{\omega }\right)^2},
\end{align}
where the approximation is valid for $k_B T \gg \tilde \omega $. The above results assume that the state close to the critical point has been obtained directly, for example, as a consequence of thermalization. Alternatively, it is possible to create the state close to the critical point by adiabatically ramping the control parameter from $g=0$ to a value close to the critical value $g_c$. { In this case, because of different energy level structure,} $\tilde \omega$ in Eq.~\eqref{eq:toymodelqfi1} has to be replaced by $\omega$, and $g$ in Eq.~\eqref{eq:toymodelqfi2}  has to be replaced by 0.

Although counter-intuitive, this increase of the quantum Fisher information with temperature can be easily explained in hindsight: the energy levels and thus the gap between them depend on the parameters of the system. For a fixed temperature, the occupation of the energy levels will strongly depend on the parameters of the system as well. The larger the ratio between temperature and gap ${k_B T}/{\tilde\omega}$, the more information about the parameters of the system will be stored in the occupation of the levels. In particular, close to the critical point of the phase transition where the energy gap vanishes, the occupations for a fixed temperature will be extremely sensitive to the parameters of the system. { On the other hand, the quantum Fisher information is larger for higher lying excited states because they are more non-classical [see Eq.~\eqref{eq:qfi0}]. Clearly the $n$th squeezed fock state is more non-classical than the squeezed vacuum state.} Therefore, the quantum Fisher information grows with the temperature. These results and explanations will serve us as a benchmark for more realistic and finite-size systems in the next sections.

\emph{Lipkin-Meshkov-Glick Model}---The Lipkin-Meshkov-Glick Hamiltonian is a paradigmatic model describing interacting spins
\begin{align}\label{eq:LMG}
    \hat H = \omega \hat S_z - \frac{g}{N}\hat S_x^2
\end{align}
and its thermodynamic limit is the toy-model Hamiltonian from the previous section~\cite{[{This can be easily seen by applying the Holstein-Primakoff transformation and $N\gg1$ approximation}]HP}. In the Hamiltonian~\eqref{eq:LMG} the $\hat S_i$ with $i \in \lbrace x, y, z \rbrace$ are the collective spin operators of $N$ spins (more generally two-level systems), $\omega$ is the energy gap between the spin levels, and $g$ is the interaction strength. A general analytic form of the eigenstates for the Lipkin-Meshkov-Glick Hamiltonian is not known. Therefore, in order to calculate the quantum Fisher information, we numerically find the thermal state close to the critical point (on both sides of the transition) and use the relation between the fidelity $F(\hat \rho, \hat \sigma) = \Big[\mathrm{tr}\sqrt{\sqrt{\hat \rho}\hat \sigma \sqrt{\hat \rho}}\Big]^2$ and the quantum Fisher information~\cite{LIU2014fidelityFisher}
\begin{align}\label{eq:fidfish}
    \mathcal{I}_\omega = 8\lim_{\delta\omega \rightarrow 0}\frac{ 1 - F\left[\hat \rho(\omega-\delta\omega/2),\hat\rho(\omega+\delta\omega/2)\right]}{(\delta \omega)^2}.
\end{align}

\begin{figure}
    \centering
    \includegraphics[width=\columnwidth]{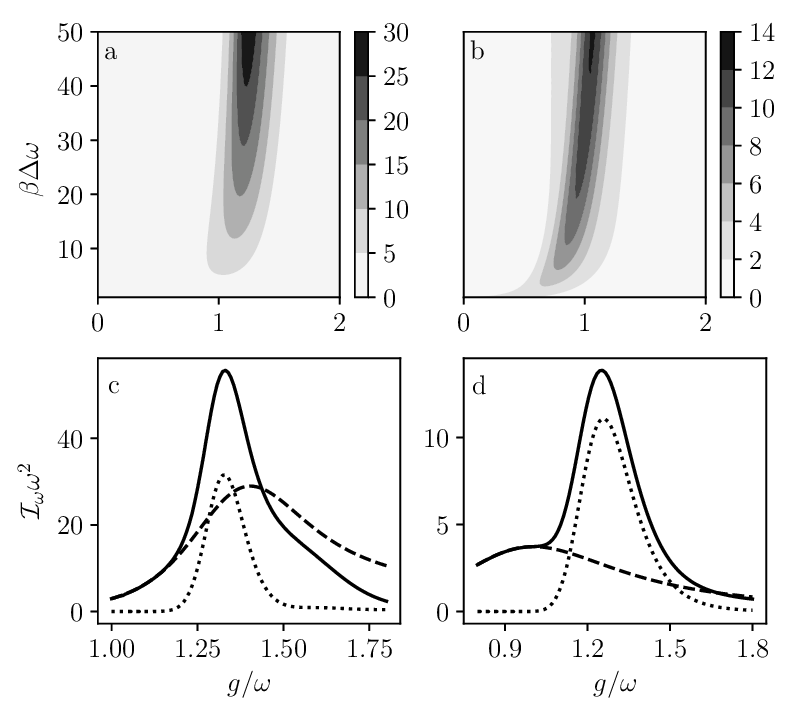}
    \caption{Top panel: quantum Fisher information as a function of temperature ($\beta =1/k_BT$) to ground state energy gap ($\Delta \omega = E_1-E_0$) ratio and interaction strength $g/\omega$ for the Lipkin-Meshkov-Glick model (a) and the Ising model (b). Bottom panel: quantum Fisher information for $\beta \Delta \omega = \infty$ (dashed line), $\beta \Delta \omega = 180$ (solid line), and the probability contribution to the quantum Fisher information (dotted line) for the Lipkin-Mehskov-Glick model (c) and the Ising model (d).}
    \label{fig:fig1}
\end{figure}

The results of our numerical simulations for $N = 20$ spins are presented in Fig.~\ref{fig:fig1}a. Note that for a finite size, the critical point of the Lipkin-Meshkov-Glick model $g_c=\omega$ is shifted towards larger values.  In agreement with our above toy-model results, the quantum Fisher information increases with temperature. However, once the temperature is too large, the quantum Fisher information falls off. We attribute this to the finite size of the system which prevents closing the energy gap completely. { In the thermodynamic limit, there is infinitely many states with quantum Fisher information growing with the excitation level $n$. In the case of finite size systems, only lower lying excited states can increase the quantum Fisher information. Exciting the state further leads to the occupation of less non-classical states for which the quantum Fisher information does not grow with $n$ anymore. This can be intuitively understood by examining the Bloch sphere picture. For an infinite Bloch sphere (thermodynamic limit toy-model), the fock states have larger radii with increasing $n$. For a finite Bloch sphere, the fock states radii increase until the equator of the Bloch sphere is reached (maximal radius and maximal entanglement) and subsequently decrease. In particular, the maximally excited state is just a collective spin up state which is not entangled.}

Interestingly, close to the critical point, the increase of the quantum Fisher information occurs for very small temperatures with respect to the ground state energy gap $\beta \Delta \omega$ for which the state is almost pure. Therefore, the origin of this increase has to be related to the information stored in the probabilities, which is absent for $k_BT = 0$, as underpinned by numerical calculations (see Fig.~\ref{fig:fig1}c).

\emph{Ising Model}---In order to make the result even more convincing, we show how the finite temperature affects the quantum Fisher information for the Ising model
\begin{align}
    \hat H =  \sum_n\left(\omega \hat \sigma_z^n - g \hat \sigma_x^n \hat \sigma_x^{n+1}\right),
\end{align}
which describes spins in a transverse magnetic field $\omega$ interacting with their respective nearest neighbors with strength $g$. The critical point of the Ising model is again $g = g_c = \omega$. In order to calculate the quantum Fisher information, we use Eq.~\eqref{eq:fidfish}. The results of our numerical simulations for a closed chain (periodic boundary conditions) for $N=6$ are presented in Fig.~\ref{fig:fig1}b and Fig.~\ref{fig:fig1}d. In spite of quantitative differences the results resemble those obtained for the Lipkin-Meshkov-Glick Hamiltonian and confirm that quantum Fisher information can benefit from a non-zero temperature. 

{ Note that, in principle, one could find the eigenspectrum of the Ising model using the Jordan-Wigner transformation and calculate the quantum Fisher information. Due to anharmonic level structure, however, it is not clear whether the quantum Fisher information could be calculated analytically for the thermal state.}
\begin{figure}
    \centering
    \includegraphics[width=\columnwidth]{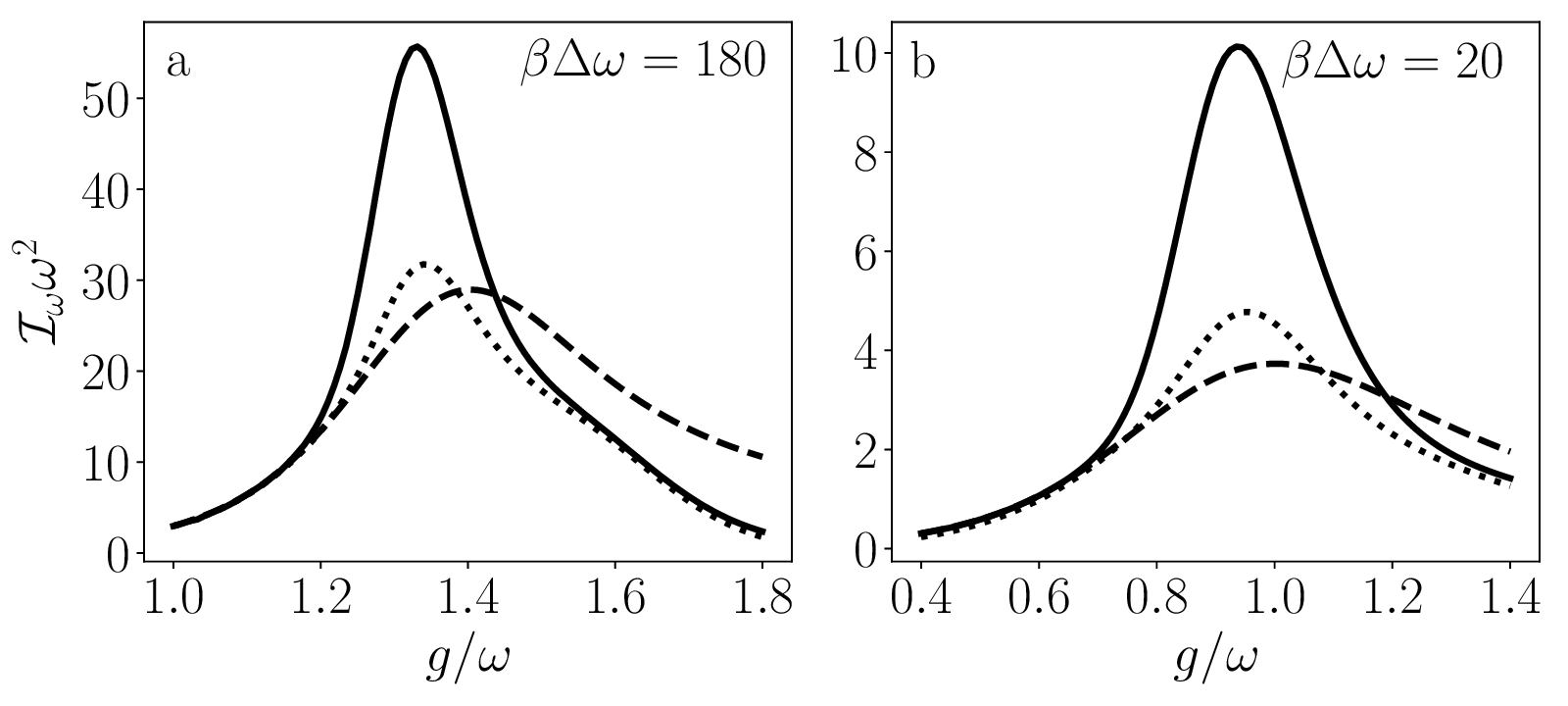}
    \caption{Numerically calculated quantum and classical Fisher information as a function of $g/\omega$. For the Lipkin-Meshkov-Glick model (a) and the Ising model (b), the classical Fisher information (dotted line) at finite temperature can be larger than the quantum Fisher information at $T=0$ (dashed line). Note that measuring $\hat S_x^2$ is not the optimal measurement choice because at a given temperature, the classical Fisher information is smaller than the quantum Fisher information (solid line).}
    \label{fig:fig2}
\end{figure}

\emph{Optimal Measurements}---In previous sections we have considered the quantum Fisher information only, which is the maximal value of the Fisher information obtained for the optimal measurement. While the quantum Fisher information might be sensitive to temperature, it is not clear whether finite temperature can indeed lead to an increased sensitivity when a particular (and simple) measurement is considered. In principle, one could perform quantum state tomography and use a maximum likelihood estimation but such an approach is not practical as it is extremely costly.

We will make an attempt to briefly address the issue of measurements in the following: for the toy model, the thermal state is an incoherent mixture of squeezed Fock states. The ground state is fully characterized by its width. Therefore, we expect that a simple strategy capable of partially characterizing an incoherent mixture of squeezed Fock states is measuring the second moment of $\hat a^\dagger + \hat a$. i.e.\
\begin{align}
 \mathrm{tr}\left[\hat \rho (\hat a^\dagger + \hat a)^2 \right] = e^{-2 \xi} \coth \left(\frac{\beta  \tilde \omega}{2}   \right),
\end{align}
whose variance is
\begin{align}
    \Delta^2\left[(\hat a^\dagger +\hat a)^2\right] = 2 e^{-4 \xi} \coth ^2\left(\frac{\beta \tilde \omega }{2} \right).
\end{align}
Using the error propagation formula, the Fisher information $\mathcal{F}_\omega$ for the measurement of $(\hat a + \hat a^\dagger)^2$ becomes
\begin{align}
   \mathcal{F}_\omega = 2\left[ \partial_\omega \xi \right]^2\left[{\beta \tilde \omega } \left(\frac{2 \omega}{g}-1 \right) \text{csch}\left(\beta  \tilde\omega  \right)+1\right]^2,
\end{align}
which, although not optimal, is always larger than the quantum Fisher information for $T=0$.

For the Lipkin-Meshkov-Glick and the Ising model, \emph{measuring the width} of the wavefunction corresponds to measuring the second moment of $\hat S_x$ operator~\cite{HP} which might be associated with spin squeezing~\cite{MA201189spins}. In Fig.~\ref{fig:fig2}a and Fig.~\ref{fig:fig2}b, we compare the numerically obtained classical and quantum Fisher information for finite $\beta\Delta \omega$ and the quantum Fisher information for $\beta\Delta \omega = \infty$ ($T=0$) as a function of $g/g_c$ for the Lipkin-Meshkov-Glick model and the Ising model, respectively. In both cases we observe that the classical Fisher information can be larger than the maximal quantum Fisher information for $T=0$ which confirms that finite temperature can increase the sensitivity of a parameter estimation. However, for a fixed temperature $T$, the classical Fisher information is smaller than the quantum Fisher information which confirms that measuring $\hat S_x^2$ is not the optimal measurement choice.

\emph{Conclusions}---In this work we have presented how non-zero temperature can be employed to increase the quantum Fisher information in crtical quantum metrology and, in principle, increase the sensitivity of measuring physical parameters in critical systems. To this end, we have considered a toy model squeezing Hamiltonian, the Lipkin-Meshkov-Glick model and the Ising model. For all these models we have shown that the quantum Fisher information increases with temperature. The main reason behind this enhancement is { the higher quantum Fisher information of some excited states and} the information about physical parameters stored in the occupation probabilities of the energy levels, which is non-existent in conventional metrological methods such as Ramsey interferometry. We would like to note that although the temperature can increase the Fisher information in critical quantum metrology, the Heisenberg limit is still the valid restriction. The enhancement is possible because critical metrology is not the optimal approach to metrology under idealized conditions~\cite{Gietka2021adiabaticcritical,gietka2022speedup}. 

Additionally, we have presented a relatively simple measurement capable of increasing the classical Fisher information for a finite temperature with respect to the quantum Fisher information at $T=0$. Yet, our proposed measurement does not saturate the Cram\'er-Rao bound. We postpone a detailed study of optimal measurements in critical metrology at finite temperature to a future investigation. 

Finally, let us emphasize that we are aware of the, for now, purely conceptual nature of our idea: we assume the the unknown parameter remains itself unchanged with temperature, and we neglect any other finite-temperature effects that might be present in the system, contributing to systematic or statistical errors in the measurements. Yet, we have seen that even a small finite temperature modifies the quantum Fisher information favorably, which leads us to believe that our discovery can be of actual relevance in real-world setups. { Note that our findings are also in agreement with a very recent study~\cite{garcíapintos2024estimation}.}

\begin{acknowledgments}
We would like to acknowledge discussions with Louis Garbe, Christoph Hotter and Lukas Fiderer. Simulations were performed using the open-source \textsc{QuantumOptics.jl}~\cite{KRAMER2018109} framework in \textsc{Julia}. K.G. is supported by the Lise-Meitner Fellowship M3304-N of the Austrian Science Fund (FWF).
\end{acknowledgments}

\end{document}